\documentclass[aps,pra,reprint,floatfix,showpacs]{revtex4-1}

\usepackage{amssymb}
\usepackage{amsmath}
\usepackage{amsfonts}
\usepackage{graphicx}

\newcommand{\lbrs}{\left[}
\newcommand{\rbrs}{\right]}

\renewcommand{\vec}[1]{\boldsymbol{#1}}

\newcommand{\beq}{\begin{eqnarray}}
\newcommand{\eeq}{\end{eqnarray}}
\renewcommand{\d}{{\text{d}}}

\newcommand{\Tr}[1]{{{\text{Tr}}\lbrs #1 \rbrs}}

\newcommand{\s}{\text{s}}

\newcommand{\Hxc}{\text{Hxc}}

\newcommand{\Ext}{{\text{ext}}}

\newcommand{\EHxc}{E_{\Hxc}}

\newcommand{\VKS}{v_{\s}}

\newcommand{\VExt}{v_{\Ext}}

\renewcommand{\vr}{\vec{r}}

\newcommand{\bra}[1]{\langle#1|}
\newcommand{\ket}[1]{|#1\rangle}

\begin{document}

\title{Kohn-Sham potentials in %
exact density-functional theory\\ at non-integer electron numbers}
\author{Tim Gould}\affiliation{Qld Micro- and Nanotechnology Centre, %
Griffith University, Nathan, Qld 4111, Australia}
\email{t.gould@griffith.edu.au}
\author{Julien Toulouse}\affiliation{Sorbonne Universit\'es,
UPMC Univ Paris 06, UMR 7616, Laboratoire de Chimie Th\'eorique,
F-75005 Paris, France\\
CNRS, UMR 7616, Laboratoire de Chimie Th\'eorique, F-75005 Paris, France}

\begin{abstract}
Within exact electron density-functional theory, we investigate
Kohn-Sham (KS) potentials, orbital energies, and non-interacting
kinetic energies of the fractional ions of Li, C and F. We use quantum
Monte Carlo densities as input, which are then fitted, interpolated at
non-integer electron numbers $N$, and inverted to produce accurate KS
potentials $v_\s^N(\vr)$. We study the dependence of the KS potential
on $N$, and in particular we numerically reproduce 
the theoretically predicted spatially constant discontinuity of
$v_\s^N(\vr)$ as $N$ passes through an integer.
We further show that, for all the cases considered, the inner orbital
energies and the non-interacting kinetic energy are nearly piecewise
linear functions of $N$. This leads us to propose a simple
approximation of the KS potential $\VKS^N (\vr)$ at any fractional
electron number $N$ which uses only quantities of the systems with the
adjacent integer electron numbers.
\end{abstract}

\maketitle

Over the past few decades, Kohn-Sham (KS)~\cite{KohnSham}
density-functional theory (DFT)~\cite{HohenbergKohn}
has become one of the most important tools in electronic-structure
theory. Given the overwhelming popularity of density-functional
approximations (DFAs) (e.g., PBE~\cite{GGA}, hybrids~\cite{Becke1993}),
surprisingly few studies have been dedicated to the detailed
properties of the \emph{exact} KS system. This is despite the fact that
unusual properties~\cite{PerdewDissociate} of the fictious
non-interacting KS system serve a vitally important role in
reproducing the quantum mechanical properties of the interacting
system in cases where degeneracies are present in the ground state,
or where electrons are added and removed.
In this paper we study exact KS DFT properties
of difficult, open quantum systems with degenerate ground states
-- specifically open-shell atoms with non-integer electron numbers.
A primary aim is to provide guidance for the construction of new DFAs.

In quantum mechanics, open electronic systems with a non-integer
average number of electrons naturally arise, for example, as fragments
from a molecular dissociation in entangled quantum states. In
particular, in DFT the study of systems with
fractional electron numbers is of great importance for a better
understanding of the theory (for a recent review, see
Ref.~\onlinecite{CohMorYan-CR-12}). For such fractional systems,
Perdew {\it et al.}~\cite{Perdew1982} proved\footnote{At least under
certain assumptions about orbital continuity.} that the energy is a
piecewise linear function of the electron number between the adjacent
integers. This lead to the theoretical prediction of the discontinuity
of the KS potential as the electron number passes through
an integer, with many important physical consequences concerning the
description of the fundamental gap~\cite{PerLev-PRL-83,ShaSch-PRL-83},
molecular dissociation~\cite{Perdew1982} or charge-transfer
excitations~\cite{Maitra2005}. This also lead to the explanation that
the underestimation of energies obtained with the usual semilocal
DFAs for delocalised densities is a
consequence of their deviations from the exact piecewise linear
behavior of the
energy~\cite{RuzPerCsoVydScu-JCP-06,MoriSanchez2006,%
RuzPerCsoVydScu-JCP-07,Cohen2008,MorCohYan-PRL-08}. These
understandings have guided the design of improved DFT
approximations~\cite{Vydrov2007,Cohen2007,MoriSanchez2009,%
Johnson2011,Gould2013-LEXX,Kraisler2013}.

Although the piecewise linear energy curve and the discontinuity of
the KS potential have been widely discussed in the literature,
little else is known about the form of the exact
KS potential as a function of the fractional electron number. In this
paper, we fill this gap by providing and analysing accurate KS
potentials as a function of the electron number for the fractional
ions of Li, C and F, exploring a range from 2 to 10 electrons. For all
these systems, we numerically reproduce the discontinuity of the KS
potential. We then propose a simple approximation of the KS potential
at any fractional electron number which uses only quantities of the
systems with the adjacent integer electron numbers.

\vskip 2mm \noindent {\it KS DFT for fractional electron numbers.} \hskip 2mm
The ground-state energy of a system with a fractional number of
electrons $N=M+f$ (where $M$ is an integer and $0 \leq f \leq 1$) can
be defined in the zero-temperature grand-canonical ensemble formalism
as~\cite{Perdew1982,Cha-JCP-99,Kraisler2013}
(see also Refs.~\onlinecite{Yang2000,Sav-CP-09} for an alternative view)
\begin{equation}
E^{N} = \min_{\hat{\Gamma}\to N} \Tr{\hat{\Gamma}
  \left( \hat{T} + \hat{V}_{\Ext} + \hat{W}_{\text{ee}} \right) },
\label{eq:Ef}
\end{equation}
where $\text{Tr}$ denotes the trace and the search is over all
normalized ensemble fermionic density matrices $\hat{\Gamma}$ yielding
$N$ electrons, i.e. $\text{Tr} [\hat{\Gamma} \hat{N}] = N$ where
$\hat{N}$ is the number operator. In Eq.~\eqref{eq:Ef}, $\hat{T}$ is
the kinetic energy operator, $\hat{V}_{\Ext}$ is the external
potential operator, and $\hat{W}_{\text{ee}}$ is the electron-electron
interaction operator. With the usual assumption that the ground-state
energy for integer electron numbers in a fixed external potential is a
convex function, $E^{M} \leq (E^{M+1} + E^{M-1})/2$, the minimizing
density matrix is obtained as a linear interpolation between
ground-state density matrices of the $M$ and ($M+1$)-electron systems
\begin{equation}
\hat{\Gamma}^{N} = (1-f) \hat{\Gamma}^{M} + f \hat{\Gamma}^{M+1},
\label{eq:Gammaf}
\end{equation}
where $\hat{\Gamma}^{M} = \sum_e w_e \ket{\Psi^M_{e}}\bra{\Psi^M_{e}}$
(with $\sum_e w_e=1$) is made of the possibly degenerate $M$-electron
ground states, and similarly for
$\hat{\Gamma}^{M+1}$. Equation~\eqref{eq:Gammaf} immediately implies
that the ground-state energy and the one-electron density are also
piecewise linear in $f$
\begin{align}
E^{N} =& (1-f) E^{M}+ f E^{M+1},
\\
n^{N}(\vr) =& (1-f) n^{M}(\vr)+ f n^{M+1}(\vr),
\label{eqn:nf}
\end{align}
where $n^{M}(\vr)$ is the density corresponding to the density matrix
$\hat{\Gamma}^{M}$, and similarly for $n^{M+1}(\vr)$. The freedom in
the choice of the ensemble weights $w_e$ can be used to impose
symmetries. For example, for atoms with an incomplete p shell, one can
choose $w_e$ so as to obtain a spherically symmetric density
$n^{M}(r)$.

In KS DFT, an universal functional can be defined for any fractional
electron density $n(\vr)$ with the constrained-search
formalism~\cite{Perdew1982}
\begin{equation}
F[n] = \min_{\hat{\Gamma}\to n} \Tr{\hat{\Gamma}
  \left( \hat{T} + \hat{W}_{\text{ee}} \right) },
\end{equation}
where $\hat{\Gamma}\to n$ refers to all normalized ensemble fermionic
density matrices $\hat{\Gamma}$ yielding the density $n(\vr)$,
i.e. $\text{Tr} [\hat{\Gamma} \hat{n}(\vr)] = n(\vr)$ where
$\hat{n}(\vr)$ is the density operator. Decomposing this functional as
\begin{equation}
F[n] = T_\s[n] + E_\Hxc[n],
\end{equation}
where $T_\s[n] = \min_{\hat{\Gamma}\to n} \text{Tr} [\hat{\Gamma}
  \hat{T}]$ is the KS non-interacting kinetic-energy functional and
$E_\Hxc[n]$ is the remaining Hartree-exchange-correlation functional,
leads to the KS expression for the exact ground-state energy
\begin{equation}
E^{N} = \min_{\hat{\Gamma}} \Bigl\{ \Tr{\hat{\Gamma}
  \left( \hat{T} + \hat{V}_{\Ext} \right)} +E_\Hxc[n_{\hat{\Gamma}}] \Bigl\},
\end{equation}
where the functional $E_\Hxc$ is evaluated at the density coming from
$\hat{\Gamma}$. The minimizing density matrix is assumed to have the
form
\begin{equation}
\hat{\Gamma}^N_{\s} = (1-f) \hat{\Gamma}_{\s}^{M,f} + f \hat{\Gamma}_{\s}^{M+1,f},
\end{equation}
where $\hat{\Gamma}_{\s}^{M,f} = \sum_e w_{e}
\ket{\Phi_{e}^{M,f}}\bra{\Phi_{e}^{M,f}}$ is a density matrix made of
$M$-electron single-determinant wave functions (which depend on $f$),
and similarly for $\hat{\Gamma}_{\s}^{M+1,f}$, and again the ensemble
weights $w_{e}$ can be used to impose symmetries. All the $M$- and
($M+1$)-electron single determinants are constructed from a common set
of KS orbitals $\{ \phi_i^N(\vr) \}$ determined by (in atomic units)
\begin{equation}
\left[-\tfrac12 \nabla^2+\VKS^{N}(\vr)\right]\phi_i^N(\vr)
=\varepsilon_i^N\phi_i^N(\vr),
\label{eqn:KSEqn}
\end{equation}
where $\VKS^{N}(\vr)$ is the KS potential,
\begin{align}
\VKS^{N}(\vr) = \VExt(\vr)+\frac{\delta\EHxc[n^{N}]}{\delta n(\vr)},
\label{eq:VKSf}
\end{align}
and $\varepsilon_{i}^N$ are the KS orbital energies
($\varepsilon_i^N\leq\varepsilon_j^N$ for $i<j$).
In Eq.~\eqref{eq:VKSf}, the KS potential is evaluated at the exact
density $n^{N}(\vr)$ which is also the density given by the KS density
matrix $\hat{\Gamma}_{\s}^N$ and can be written in terms of the KS
orbitals and fractional occupation numbers $\{ f_i \}$
\begin{equation}
n^{N}(\vr) = \sum_i f_i |\phi_i^N(\vr)|^2,
\label{eqn:nKS}
\end{equation}
where $f_i = 2$ for the $N_{\text{inner}}$ lowest inner doubly
occupied orbitals, $\sum_h f_h = N-2N_{\text{inner}}$ for the
fractionally occupied degenerate highest occupied orbitals (HOMOs),
and $f_i = 0$ for the
remaining unoccupied orbitals. Unlike other
work~\cite{GriBae-JCP-04,BogJacRei-JCP-13}, we use a spin-restricted
open-shell formalism, and thus avoid dealing with spin dependence.

\begin{figure*}[bt]
\includegraphics[width=0.33\linewidth]{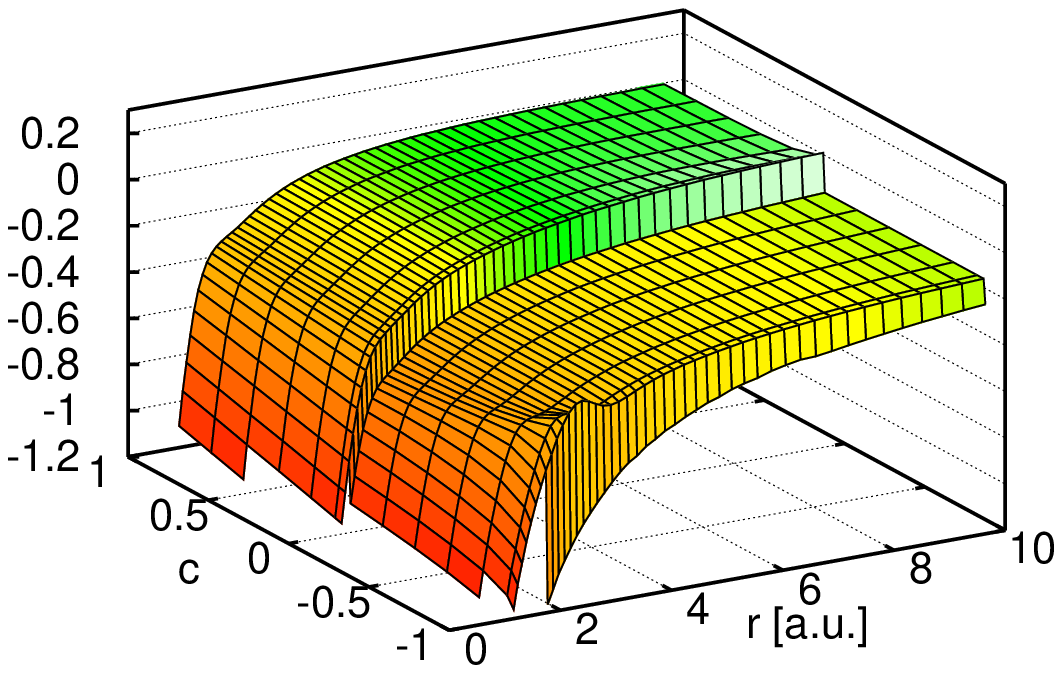}%
\includegraphics[width=0.33\linewidth]{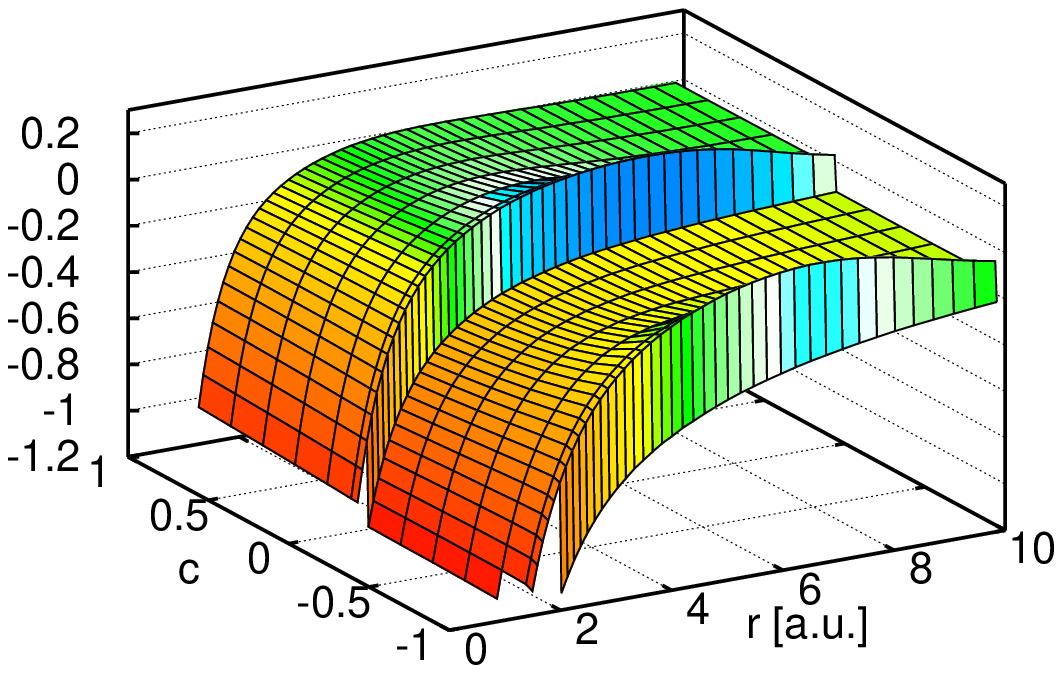}%
\includegraphics[width=0.33\linewidth]{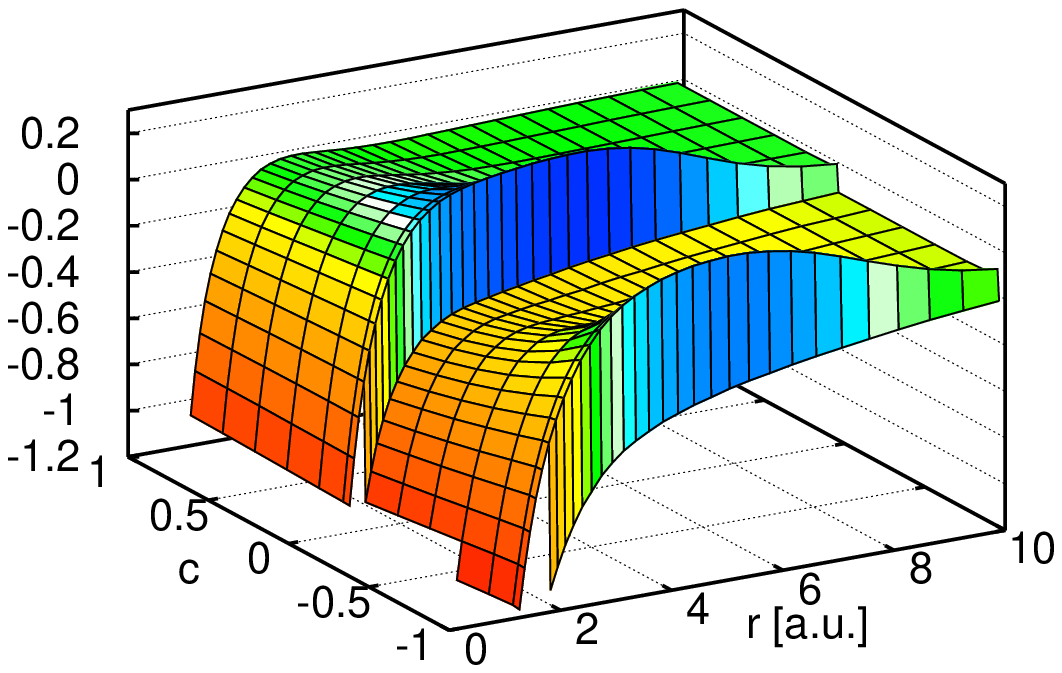}%
\caption{(Color online)
  KS potentials $\VKS^{Z+c}(r)$ [hartree] as a function of the
  radial distance $r$ and the fractional excess electron number
  $c=N-Z$ for Li (left), C (middle) and F (right).
  Colors indicate height (from red to green for $\VKS^{Z+c}(r)<0$, white for $\VKS^{Z+c}(r)\approx0$, blue for $\VKS^{Z+c}(r)>0$).
\label{fig:VKS}}
\end{figure*}
In principle, if density variations changing the electron number are
allowed in the functional derivative of Eq.~\eqref{eq:VKSf}, the KS
potential $\VKS^{N}(\vr)$ including its spatial-constant component is
uniquely determined. In practice, for any fixed $N$, if an accurate
electron density $n^{N}(\vr)$ is known, one can find the corresponding
KS potential $\VKS^{N}(\vr)$ obeying
$\VKS^{N}(|\vr|\to\infty)=0$ by ensuring that the orbitals satisfying
Eq.~\eqref{eqn:KSEqn} give the density through Eq.~\eqref{eqn:nKS}. In
this work, we make use of this mapping to obtain accurate KS
potentials $\VKS^{N}(\vr)$, orbital energies $\varepsilon_{i}^N$, and
KS kinetic energies $T_s^N \equiv T_\s[n^{N}]$ as a function of
the fractional electron number $N$.

\vskip 2mm \noindent {\it Computational method.} \hskip 2mm
We consider atoms with external potential
$v_{\text{ext}}(r)=-Z/r$. Using Eq.~\eqref{eqn:nf}, the density for an
ion with any fractional electron number between $Z-1$ and $Z+1$ can be
obtained from the densities for $Z-1$, $Z$, and $Z+1$ electrons. We
thus first calculate (spherically averaged) accurate densities
$n^{M}(r)$ for these integer electron numbers using quantum Monte
Carlo (QMC). Specifically, the densities are calculated in diffusion
Monte Carlo with an improved statistical
estimator~\cite{TouAssUmr-JCP-07}, using Jastrow full-valence
complete-active-space wave functions fully optimized at the
variational Monte Carlo level~\cite{TouUmr-JCP-08}. As an accurate
inversion requires a density to be free from statistical errors, we
fit these QMC densities with a simple, yet asymptotically accurate
(as $r\to\infty$) function
that ensures that the density corresponds to a HOMO
with the appropriate eigenenergy $\varepsilon_h$,
and to a KS potential with correct asymptotic behavior
$\VKS^{M}(r\to\infty)\sim (-Z+M-1)/r$.
The fitting function and parameters given in the supplementary
material~\cite{Supp} give densities that are accurate to within
the QMC statistical error.

After obtaining the density at fractional electron numbers $n^{N}(r)$
via Eq.~\eqref{eqn:nf}, we calculate the KS potential by using a
numerically stabilised modification of Wang and Parr's iterative
approach~\cite{Wang1993}. The KS potential at iteration $m+1$ is found
from the quantities at the previous iteration $m$ through
$\VKS^{N,m+1}(r)=\VKS^{N,m}(r) + Q\frac{n^{N}(r)-n^{N,m}(r)}{G^{N,m}(r)}$,
where $G^{N,m}(r)=\sum_i f_i|\phi_{m}^{N,m}(r)|^2/\varepsilon_{i}^{N,m}
+ \max[0,(1/\varepsilon_h-1/\varepsilon_{h}^{N,m})\{n^{N}(r)-n^{N,m}(r)\}]$,
and $Q>0$ is a convergence parameter.
By starting from fractional LEXX~\cite{Gould2013-LEXX}
potentials and orbitals we achieve
$\int\d\vr |n^{N}(\vr)-n^{N,m}(\vr)|<10^{-6}$,
albeit with increasing errors in $\VKS^N$ for $N\leq Z-0.8$.

\begin{figure}
\includegraphics[width=0.9\linewidth]{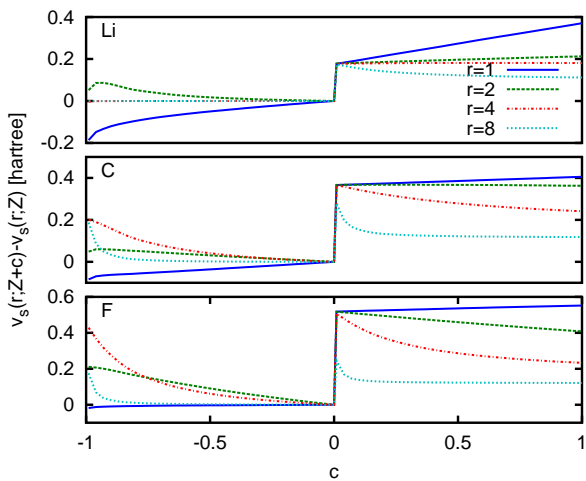}%
\caption{(Color online)
  Change in Kohn-Sham potentials $\VKS^{Z+c}(r)-\VKS^{Z^-}(r)$ as a
  function of $c>-1$ for Li (top), C (middle) and F (bottom), for
  selected values of $r$.
  \label{fig:DVKS}}
\end{figure}

\vskip 2mm \noindent {\it Results and discussion.} \hskip 2mm
In Figure~\ref{fig:VKS} we show the KS potentials $\VKS^{Z+c}(r)$
(plots of its components are shown in the supplementary material~\cite{Supp})
as a function of the radial distance $r$ and the fractional excess
electron number $c=N-Z$ where $-1\leq c\leq 1$, i.e. for ions from
A$^+$ through A to A$^-$ for Li, C and F. Calculations are performed
for $c\in \{-1,-0.99,-0.95,-0.8,-0.6,-0.4,-0.2,0,%
0.01,0.05, \allowbreak0.2,0.4,0.6,0.8,1\}$.

The most obvious feature of these plots is the presence of a
discontinuity in the potentials as the electron number $N$ crosses
an integer. Let us concentrate on the discontinuity at
$N=Z$, i.e. $c=0$. At first sight, in Figure~\ref{fig:VKS}, this
discontinuity seems to be dependent on $r$, decreasing at large
distances. To analyse the discontinuity more precisely we plot
in Figure~\ref{fig:DVKS} the quantity 
$\VKS^{Z+c}(r) - \VKS^{Z^-}(r)$ as a function of $c$ (interpolated in
$-0.99\leq c\leq 0$ and $0.01\leq c\leq 1$)
for selected values of $r$.
It is now clear that, if we
extrapolate the plots from $c>0$ toward the limit
$c\to 0^+$, the results are consistent with the KS potential having
a constant discontinuity
\begin{equation}
  \Delta^Z = \VKS^{Z^+}(r) - \VKS^{Z^-}(r) = I-A,
  \label{eqn:Delta}
\end{equation}
at $N=Z$, independent of $r$. For the open-shell systems considered here, the discontinuity
$\Delta^Z$ is equal to the difference between the ionization energy $I$ 
and the electronic affinity $A$ of the neutral atom.
Therefore, the present results numerically reproduce the 
theoretically predicted~\cite{GoriGiorgi2009,Perdew2009Limits}
spatially constant discontinuity of the KS potential when the electron
number $N$ crosses an integer $M=3$, $6$, or $9$, which had been so
far numerically observed only for
$M=1$~\cite{SagPer-PRA-08,Perdew2009Limits,GoriGiorgi2009}. 
The values of $\Delta^Z$ are found to be $0.178$, $0.368$ and $0.519$ hartree,
for Li, C and F, respectively, very close to
$0.175$, $0.366$ and $0.516$ hartree found using $I-A$.

Beside the discontinuity, it is interesting to note that
the potential is sensitive to the order of the
limits $c\to 0^+$ and $r\to\infty$~\cite{Perdew2009Limits}.
It is clear from Figures~\ref{fig:VKS} and \ref{fig:DVKS} that
as $r$ increases the KS potentials go to zero in a larger and
larger range of $c$, except near the limit $c\to 0^+$ where it
becomes more and more curved. This is understandable since
$\VKS^{Z+c}(r) - \VKS^{Z^-}(r)$ must go to $\Delta^Z$ in the limit
$c\to0^+$, but at the same time, for any finite 
value of $c$, the KS potential for $M < N \leq M+1$ must have the same
asymptotic behavior as the KS potential for the $(M+1)$-electron
system, $\VKS^{N}(r\to\infty) \sim \VKS^{(M+1)^-}(r) \sim
(-Z+M)/r$~\footnote{The asymptotic behavior of the potential is
determined by the asymptotic behavior of the ensemble density in
Eq.~\eqref{eqn:nf} which is dominated for large $r$ by the
$(M+1)$-electron density}, and therefore must go to zero for large
$r$.

\begin{figure}
\includegraphics[width=0.9\linewidth]{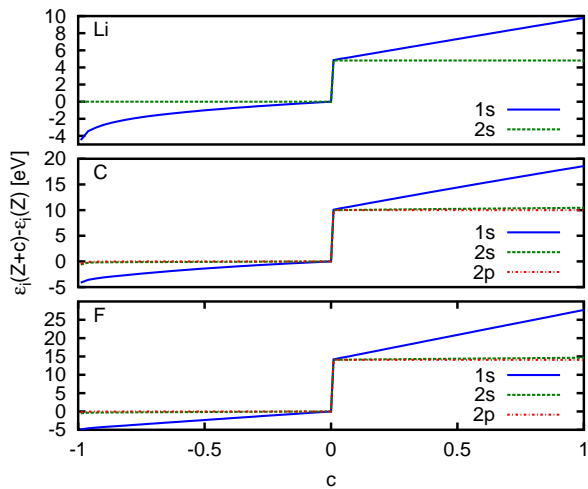}%
\caption{(Color online)
  Change in Kohn-Sham orbital energies for all orbitals
  $\varepsilon_i^{Z+c}-\varepsilon_i^{Z^-}$ as a function of
  $c>-1$ for Li (top), C (middle) and F (bottom).
  \label{fig:Eigs}}
\end{figure}
\begin{figure}
  \includegraphics[width=0.9\linewidth]{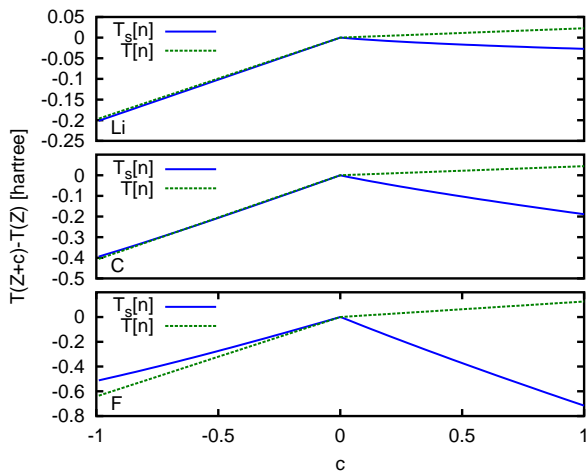}%
  \caption{(Color online)
    Change in Kohn-Sham kinetic energy $T_s^{Z+c}-T_s^Z$ and
    kinetic energy $T^{Z+c}-T^Z$ as a function of $c$.
    \label{fig:KSEns}}
\end{figure}

In Figure~\ref{fig:Eigs} we plot as a function of $c$ the difference
between the orbital energies at the fractional electron number $Z+c$
and their neutral atomic values:
$\varepsilon_i^{Z+c}-\varepsilon_i^{Z^-}$.
As expected, all orbital energies have a discontinuity of $\Delta^Z$
at $c=0$. As required by theory the HOMO energy is constant,
within numerical noise, between integer electron numbers:
$\varepsilon_{h}^N=\varepsilon_{h}^{M^-}+\Delta^M=\varepsilon_{h}^{(M+1)^-}$.
Interestingly, we find that the energies of the inner orbitals 
follow an almost piecewise linear behaviour
\begin{equation}
  \varepsilon_{i}^N \approx (1-f) [ \varepsilon_{i}^{M^-} +
    \Delta^M ]
  + f \varepsilon_{i}^{(M+1)^-},
  \label{eqn:epsinner}
\end{equation}
where $i \leq N_{\text{inner}}$.
As far as we know, this is the first time that the energies of inner
orbitals are studied as a function of the fractional electron number.
We also plot in Figure~\ref{fig:KSEns} the KS kinetic energy $T_s^{Z+c}$
and the exact kinetic energy $T^{Z+c}=-E^{Z+c}$ (by the virial
theorem) as a function of $c$, adjusted by their neutral atomic $c=0$
values. Unlike $T$, $T_s$ is not piecewise linear but was
theoretically shown to be a convex function of
$f$~\cite{Levy2014}. Our numerical results confirm this convex
behavior. However, for all the systems studied here, $T_s$ turns out
to be remarkably close to linearity
\begin{equation}
  T_s^N \approx (1-f) T_s^M + f T_s^{M+1},
  \label{eqn:ts}
\end{equation}
with the maximum deviation from linearity being $-3.3$ mhartree for
Li$^{-0.4}$, $-8.9$ mhartree for C$^{-0.4}$ and $-15.3$ mhartree for
F$^{+0.6}$.
This near linearity was previously observed only for
weakly interacting two-electron systems~\cite{GorSav-JPCS-08}.

The near linearity of $T_s^N$ and $\varepsilon_i^N$ suggest that
the KS orbitals change in a predictable fashion as partial electrons
are added to the atoms, even in the difficult, degenerate systems
considered here. If Eqs.~\eqref{eqn:epsinner} and \eqref{eqn:ts}
apply more generally they could provide desirable local and
global constraints on  new DFAs.
Furthermore, since the KS kinetic
energy $T_s^N$, the KS potential $v_\s^N(\vr)$,
and the orbital energies $\varepsilon_i^N$ are linked
by the KS energy expression
$T_s^N + \int n^N(\vr) \VKS^N(\vr) \d \vr  = \sum_i f_i \varepsilon_i^N$
 at fractional electron number $N$,
the linear approximations of Eqs.~\eqref{eqn:epsinner}
and~\eqref{eqn:ts} suggest a linear approximation for $n^{N}(\vr)
\VKS^N (\vr)$
\begin{eqnarray}
  n^{N}(\vr) \VKS^N (\vr) &\approx&
  (1-f)n^{M}(\vr)[\VKS^{M^-}(\vr)+\Delta^M ] 
  \nonumber\\
  &&+ f n^{M+1}(\vr)\VKS^{(M+1)^-}(\vr).
  \label{eqn:VKSA}
\end{eqnarray}
Indeed, we found that Eq.~\eqref{eqn:VKSA} is an excellent
approximation to $\VKS^N (\vr)$ for all cases studied here, especially when
considered against the electronic density.
Details are provided in the supplementary material~\cite{Supp}.
We note that this approximation becomes exact
in two limits: close to a nucleus where the KS potentials are
dominated by the diverging electron-nucleus potential
$v_\text{ext}(\vr)$, and in the asymptotic region $|\vr|\to\infty$
where the density $n^{M+1}(\vr)$ dominates over $n^{M}(\vr)$ and the
KS potential $\VKS^N (\vr)$ reduces to $\VKS^{(M+1)^-}(\vr)$. It can
also be shown that this approximation is exact in the trivial case
$0 \leq N \leq 1$ and in the case $1 \leq N \leq 2$ if $n^{2}(\vr)
= 2n^{1}(\vr)$.
Preliminary work
suggests that this approximation may be valid in a wider range
of systems than those explored here.

\vskip 2mm \noindent {\it Conclusions.} \hskip 2mm
We have shown that accurate KS potentials $\VKS^N (\vr)$
at non-integer electron numbers $N$
can be obtained by inversion of accurate \emph{ab initio}
electron densities. This has allowed us to numerically reproduce
on systems with more than two electrons the theoretically predicted
spatially constant discontinuity of the KS potential when the electron
number crosses an integer.
We have also found that, for all the atomic systems studied here,
both the energies $\varepsilon_i^N$ of the inner orbitals
(below the HOMO) 
and the KS kinetic energy $T_s^N$ are nearly piecewise linear
functions of $N$. This has lead us to propose the simple approximation
of Eq.~\eqref{eqn:VKSA} for the KS potential $\VKS^N (\vr)$ 
at any fractional electron number $N$ which uses only quantities of
the systems with adjacent integer electron numbers. This approximation
appears to work very well for all the cases considered and its
generality and potential application to fragment and partition
DFT~\cite{Elliott2010,Tang2012,Fabiano2014} will be explored in
future work.

\acknowledgments
TG received computing support from the
Griffith University Gowonda HPC Cluster.

%

\end{document}